\documentclass[11pt]{article}
\usepackage{amssymb,latexsym,amsmath} 
\usepackage{amsfonts, graphics}
\newcommand{\R}{\mathbb R}

\def\open#1{\setbox0=\hbox{$#1$}
\baselineskip = 0pt
\vbox{\hbox{\hspace*{0.4 \wd0}\tiny $\circ$}\hbox{$#1$}}
\baselineskip = 11pt\!}
\def\fn{\open{f}}

\def\Rho{{\cal R}}

\newcommand{\prfe}{\hspace*{\fill} $\Box$

\smallskip \noindent}

\def\be{\begin{equation}}
\def\ee{\end{equation}}
\def\bea{\begin{eqnarray}}
\def\eea{\end{eqnarray}}
\def\beas{\begin{eqnarray*}}
\def\eeas{\end{eqnarray*}}

\def\supp{\mathrm{supp}\,}

\begin{document}

\newtheorem{theorem}{Theorem}[section]
\renewcommand{\thetheorem}{\arabic{section}.\arabic{theorem}}
\newtheorem{definition}[theorem]{Definition}
\newtheorem{proposition}[theorem]{Proposition}
\newtheorem{example}[theorem]{Example}
\newtheorem{remark}[theorem]{Remark}
\newtheorem{cor}[theorem]{Corollary}
\newtheorem{lemma}[theorem]{Lemma}

\title{The formation of black holes in spherically symmetric gravitational collapse}

\author{H{\aa}kan Andr\'{e}asson\\
        Mathematical Sciences\\
        Chalmers University of Technology\\
        G\"{o}teborg University\\
        S-41296 G\"oteborg, Sweden\\
        email: hand@math.chalmers.se\\
        \ \\
        Markus Kunze\\
        Fachbereich Mathematik\\
        Universit\"at Duisburg-Essen\\
        D-45117 Essen, Germany\\
        email: markus.kunze@uni-due.de\\
        \ \\
        Gerhard Rein\\
        Mathematisches Institut der Universit\"at Bayreuth\\
        D-95440 Bayreuth, Germany\\
        email: gerhard.rein@uni-bayreuth.de}

\maketitle

\begin{abstract}
We consider the spherically symmetric, asymptotically flat
Einstein-Vlasov system. We find explicit conditions on the initial data, with ADM mass 
$M,$ such that 
the resulting spacetime has the following properties: there is a family of radially outgoing 
null geodesics where the area radius $r$ along each geodesic is bounded by $2M,$ the timelike 
lines $r=c\in [0,2M]$ are incomplete, 
and for $r>2M$ the metric converges asymptotically to the Schwarzschild metric with mass $M$. 
The initial data that we construct guarantee the formation of a black hole in the evolution. 
We also give examples of such initial data with the additional property that the solutions 
exist for all $r\geq 0$ and all Schwarzschild time, i.e., we obtain global existence in Schwarzschild coordinates in situations 
where the initial data are not small. 
Some of our results are also established for the Einstein equations coupled to a general 
matter model characterized by conditions on the matter quantities. 
\end{abstract}

\section{Introduction}
\setcounter{equation}{0}
One of the many striking predictions of General Relativity is the
assertion that under appropriate conditions astrophysical objects
like stars or galaxies undergo a gravitational collapse resulting
in a spacetime singularity. This was first proven by Oppenheimer and
Snyder \cite{OS} who constructed a semi-explicit example of a homogeneous
spherically symmetric ball of dust, i.e., of a pressure-less fluid,
which under its self-consistent, general relativistic gravitational
interaction collapses. During this collapse the scalar
curvature of spacetime blows up at the centre of symmetry,
and the geometry of spacetime breaks down there.
This is referred to as the formation of a spacetime singularity.
An important feature of the Oppenheimer-Snyder solution is that during
the collapse a two-dimensional spacelike sphere evolves which encloses
the singularity and through which no causal curve, i.e., no light
ray or particle trajectory, can pass outward. In this way the
spacetime singularity is isolated from the outside part of spacetime
by a so-called event horizon, and the singularity cannot be seen
or in any other way be experienced by observers outside the event
horizon. This configuration was later termed a black hole.

In the 1960s Penrose \cite{pen} proved that the formation of spacetime singularities
from regular initial data is not restricted to spherically symmetric,
especially constructed or isolated examples but is a genuine,
stable feature of spacetimes. However, this result gives little information
about the geometric structure of a spacetime with such a singularity.
In particular, it is in general not known if every spacetime singularity
which arises from the gravitational collapse of regular initial data
is covered by an event horizon. Since the existence of so-called naked singularities
(for which, by definition, the latter is not true) would violate predictability
(it would not be possible to predict from the initial data
what an observer would see if he could observe a singularity),
the cosmic censorship conjecture was formulated which demands that any singularity which
arises from the gravitational collapse of \textit{generic} regular initial data
is indeed hidden behind an event horizon. The restriction to generic data
means that naked singularities are allowed to occur for a ``null set''
of the initial data. An important example where naked singularities do form
for a null set, but for which cosmic censorship holds true,
is the spherically symmetric Einstein-scalar field system, cf.\ \cite{Chr94,Chr99a}.
Actually the above is an informal statement of the so-called weak cosmic censorship
conjecture \cite[12.1]{wald}; we will not be concerned with the strong version in the present paper.
For a mathematical discussion and the definition of the weak cosmic censorship
conjecture we refer to \cite{Chr99}.

To deal with this conjecture in full generality is out of reach of
the present level of mathematics, but under
the assumption of spherical symmetry progress has been made in recent years.
One important outcome of these investigations is that
the answer is sensitive to which model is chosen
to describe the matter. Christodoulou \cite{Chr84} showed that for dust,
i.e., the matter model used by Oppenheimer and Snyder, cosmic censorship
is violated. On the other hand, in a series of papers Christodoulou
investigated a massless scalar field as matter model and showed in 1999
that weak and strong cosmic censorship hold true
for this matter model; see \cite{Chr99a} and the references therein.

In the present investigation the main example considered as a matter model
is the so-called collisionless gas as described by the Vlasov equation.
It is used extensively in astrophysics, cf.\ \cite{BT}, to describe galaxies or
globular clusters which are viewed as large ensembles of mass points
which interact only through the gravitational field that the
ensemble creates collectively. In a relativistic context this leads to
the Einstein-Vlasov system. All results available for this system support the following

\smallskip

\noindent
{\bf Conjecture:} {\em Weak cosmic censorship holds for the
Einstein-Vlasov system.}

\smallskip

\noindent
We mention explicitly that, in contrast to dust,
small, spherically symmetric initial data launch global solutions,
i.e., the solutions are geodesically complete and hence
satisfy cosmic censorship, cf.\ \cite{RR1}.
Also, the numerical simulations \cite{AR1,OC,RRS2}
which treat large initial data support the hypothesis that naked singularities
do not form in the evolution. We point out a further interesting feature
of Vlasov matter observed in these numerical studies: In a one-parameter family
of solutions which for large parameters, i.e., large amplitudes of the initial data,
collapse to a black hole the smallest black hole always has a
strictly positive ADM mass, i.e., there is a mass gap. For some other models, e.g. 
a scalar field, the mass of the black hole as a function 
of the parameter is continuous and arbitrarily small black holes can form, 
cf.\ \cite{CG} for a review. 

The aim of the present paper is to find explicit conditions
on the initial data which ensure the formation of black holes.
This class of initial data has the important property that,
except for ``boundary cases'', properly restricted small perturbations
of the data lead to solutions with the same properties. In this
sense the established behaviour of the solutions is
stable and not restricted to especially constructed
solutions or initial data, respectively. It turns
out that some of our results can be formulated
for a general matter model which satisfies certain specific assumptions,
and in order to give a broader impact to our results we shall do so. 
At the same time we emphasize that the Vlasov matter model is the only one
which is presently known to actually satisfy all the assumptions needed
for our arguments to go through. 

As an interesting corallary to our main result we show that it is in fact 
possible to choose initial data for the Einstein-Vlasov system, which 
lead to formation of black holes, such that the solutions exist 
for all Schwarzschild time and all $r\geq 0.$ We thus obtain global 
existence in Schwarzschild coordinates for initial data which are not small, and  
to the best of our knowledge this is the first global existence 
result in Schwarzschild coordinates for initial data which lead to 
gravitational collapse and formation of black holes.

One aspect of our result is that there is a set
of initial data which leads to gravitational collapse
such that weak cosmic censorship holds. This point should be related
to an earlier result by Rendall \cite{Rend92}, where it is shown that there exist
initial data for the spherically symmetric Einstein-Vlasov system
such that a trapped surface forms in the evolution.
The occurrence of a trapped surface signals the formation of an event horizon.
Indeed, Dafermos \cite{D05} has proved that if a spherically symmetric spacetime
contains a trapped surface and the matter model satisfies certain hypotheses
then weak cosmic censorship holds true. In \cite{DR05} it was then
shown that Vlasov matter does satisfy the required hypotheses. Hence, by combining
these results it follows that initial data exist which lead to gravitational collapse
and for which weak cosmic censorship holds. However, the proof in \cite{Rend92}
rests on a continuity argument, and it is not possible to tell whether or not a
given initial data set will give rise to a black hole. This is in contrast
to the explicit conditions on the initial data, together with the detailed asymptotic 
structure, that we obtain in the present work. In this regard
it is natural to relate our results to those of Christodoulou
on the spherically symmetric Einstein-scalar field system \cite{Chr87} and \cite{Chr91}.
In \cite{Chr87} it is shown that if the final Bondi mass $M$ is different from zero, 
the region exterior to the sphere $r=2M$ tends to the Schwarzschild metric 
with mass $M$. 
In Theorem \ref{bh} below we show that solutions of the spherically Einstein-Vlasov system, 
under certain conditions on the initial data, also converge to the Schwarzschild metric 
asymptotically. 
Furthermore, in \cite{Chr91} explicit conditions on the initial data are specified
which guarantee the formation of trapped surfaces. This paper played a crucial role
in Christodoulou's proof \cite{Chr99a} of the weak and strong cosmic censorship conjectures
mentioned above. The conditions on the initial data in \cite{Chr91} allow the ratio 
of the Hawking mass and the area radius to cover the full range, i.e., $2m/r\in (0,1),$ 
whereas our conditions always require $2m/r$ to be quite close to one. However, 
we believe that 
to understand gravitational collapse in the case of Vlasov matter the essential situation 
is when $2m/r$ is large. 
We thus hope that the results in the present paper will lead to progress on 
the general understanding of gravitational collapse and the weak cosmic censorship conjecture 
in the case of Vlasov matter. 

The Vlasov matter model has a further property to recommend
it when compared to other matter models.
For the Vlasov-Poisson system, which arises
as the Newtonian limit of the Einstein-Vlasov system in a rigorous sense 
\cite{RR2,Rl0}, and which is used extensively in astrophysics, there is a 
global existence and uniqueness result
for general, smooth initial data \cite{LP,Pf}. This means in particular
that any breakdown of a solution of the Einstein-Vlasov system
can be expected to be a genuine, general relativistic effect
such as a spacetime singularity and not only remainder of some bad behaviour
which the matter model exhibits already on the Newtonian level.

To be more specific, consider now a smooth spacetime manifold $M$
equipped with a  spacetime metric $g_{\alpha \beta}$; Greek indices
run from $0$ to $3$. Then the Einstein equations read
\begin{equation} \label{feqgen}
   G_{\alpha \beta} = 8 \pi T_{\alpha \beta},
\end{equation}
where $G_{\alpha \beta}$ is the Einstein tensor, a non-linear second order
differential expression in the metric $g_{\alpha \beta}$, and
$T_{\alpha \beta}$ is the energy-momentum tensor given by the
matter content (or other fields) of the spacetime. To obtain a closed system,
the field equations (\ref{feqgen}) have to be supplemented by
\begin{equation} \label{matevol}
   \mbox{evolution equation(s) for the matter}
\end{equation}
and
\begin{equation} \label{emtdef}
   \mbox{the definition of $T_{\alpha \beta}$ in terms of the matter and the metric}.
\end{equation}
It is often possible to specify conditions on (\ref{matevol}) and
(\ref{emtdef}) under which one can establish geometric properties
of a spacetime described by the Einstein-matter system
(\ref{feqgen}), (\ref{matevol}), (\ref{emtdef}).
The Penrose singularity theorem mentioned above is of
this nature, and part of our arguments will also be presented
in this form.

However, in order to verify such general conditions,
in particular with respect to the existence of local or global solutions
to the corresponding initial value problem, a specific matter model
must be chosen, and in the present paper this is a collisionless gas.
All the particles in the gas are assumed to have the same
rest mass, normalized to unity, and to move forward in time.
Hence, their number density $f$ is a non-negative function
supported on the mass shell
\[
PM := \left\{ g_{\alpha \beta} p^\alpha p^\beta = -1,\ p^0 >0 \right\},
\]
a submanifold of the tangent bundle $TM$ of the spacetime manifold $M$;
$p^\alpha$ are the canonical momenta corresponding to general
coordinates $x^\alpha=(t,x^a)$ on $M$.
We use coordinates $(t,x^a)$ with zero shift, and
Latin indices run from $1$ to $3$.
On the mass shell $PM$ the variable $p^0$ becomes a function of the
remaining variables $(t,x^a,p^b)$:
\[
p^0 = \sqrt{-g^{00}} \sqrt{1+g_{ab}p^a p^b}.
\]
The number density
$f=f(t,x^a,p^b)$ satisfies a continuity equation, the so-called Vlasov equation, which
says that $f$ is constant along the geodesics of the spacetime metric,
\begin{equation} \label{vlgen}
   \partial_t f + \frac{p^a}{p^0}\,\partial_{x^a} f
   -\frac{1}{p^0}\,\Gamma^a_{\beta \gamma} p^\beta p^\gamma\,\partial_{p^a} f = 0,
\end{equation}
where $\Gamma^\alpha_{\beta \gamma}$ are the Christoffel symbols
induced by the metric $g_{\alpha \beta}$. The energy-momentum tensor
is given by
\begin{equation} \label{emtvlgen}
   T_{\alpha \beta}
   =\int p_\alpha p_\beta f \,|g|^{1/2} \,\frac{dp^1 dp^2 dp^3}{-p_0},
\end{equation}
where $|g|$ denotes the modulus of the determinant of the metric.
The system (\ref{feqgen}), (\ref{vlgen}), (\ref{emtvlgen})
is the Einstein-Vlasov system in general coordinates.
For an introduction to relativistic kinetic theory and the Einstein-Vlasov system
we refer to \cite{And05} and \cite{Rend05}.

If, for comparison, the matter is to be described as a perfect
fluid with density $\Rho$, four-velocity field $U^\alpha$, and
pressure $P$, then the matter evolution equations are the Euler
equations
\[
U^\alpha \nabla_\alpha \Rho + (\Rho +P) \nabla^\alpha U_\alpha = 0,
\]
\[
(\Rho +P)U^\alpha \nabla_\alpha U_\beta +
(g_{\alpha \beta} + U_\alpha U_\beta) \nabla^\alpha P = 0,
\]
where $\nabla_\alpha$ is the covariant derivative corresponding to
the metric $g_{\alpha \beta}$.
The energy-momentum tensor in this case is
\[
T_{\alpha \beta}
= \Rho U_\alpha U_\beta + P (g_{\alpha \beta} + U_\alpha U_\beta ).
\]
To close the Einstein-Euler system it has to be supplemented by an equation
of state $P=P(\Rho)$. The choice $P=0$ yields the dust
matter model referred to above.

Due to the complexity of the field equations (\ref{feqgen})
very little can be said about the questions at hand for these
equations in their general form. Since on the other hand these questions
are of considerable interest also in spacetimes satisfying
simplifying symmetry assumptions, we from now on focus
on asymptotically flat, spherically symmetric spacetimes
and write down the metric
\[
ds^2=-e^{2\mu(t,r)}dt^2+e^{2\lambda(t,r)}dr^2+r^2(d\theta^2+\sin^2\theta\,d\varphi^2)
\]
in Schwarzschild coordinates. Here $t\in\R$ is the time coordinate,
$r\in [0, \infty[$ is the area radius,
i.e., $4 \pi r^2$ is the area of the orbit
of the symmetry group $\mathrm{SO}(3)$ labeled by $r$, and
the angles $\theta\in[0, \pi]$ and $\varphi\in[0, 2\pi]$
parameterize these orbits.
Asymptotic flatness means that the metric quantities $\lambda$ and $\mu$
have to satisfy the boundary conditions
\begin{equation}\label{boundc}
   \lim_{r\to\infty}\lambda(t, r)=\lim_{r\to\infty}\mu(t, r)=0.
\end{equation}
For a metric of this form the $00$, $11$, and $01$ components
of the Einstein equations are found to be
\begin{equation}\label{ein1}
   e^{-2\lambda}(2r\lambda_r-1)+1=8\pi r^2 e^{-2\mu} T_{00},
\end{equation}
\begin{equation} \label{ein2}
   e^{-2\lambda}(2r\mu_r+1)-1 = 8\pi r^2 e^{-2\lambda} T_{11},
\end{equation}
\begin{equation} \label{ein3}
   \lambda_t = 4 \pi r T_{01},
\end{equation}
where subscripts indicate partial derivatives.
The $22$ and $33$ components are also nontrivial,
but they are not needed for our
analysis, and the remaining components vanish identically due to the
symmetry assumption.

Our aim is to find explicit conditions on the initial data 
such that the corresponding solutions
of the spherically symmetric, asymptotically flat version of the
system (\ref{feqgen}), (\ref{matevol}), (\ref{emtdef}) have the following
property:
There is an outgoing radial null geodesic
$\gamma^+$ originating from $r=r_0>0$, i.e.,
\begin{equation}\label{gamma+}
   \frac{d \gamma^+}{ds}(s)=e^{(\mu-\lambda)(s,\gamma^+(s))},\;\gamma^+(0)=r_0,
\end{equation}
such that the solution exists on the outer region
\begin{equation} \label{ddef}
   D:=\{(t,r) \in [0,\infty[^2 \mid r \geq \gamma^+(t)\},
\end{equation}
and $\gamma^+$ has the property that
\begin{equation}\label{limgamma}
   \lim_{s\to \infty}\gamma^+(s) < \infty.
\end{equation}
This indicates that the matter distribution 
undergoes a gravitational collapse, and a black hole forms.
In the case of Vlasov matter we obtain a more detailed
picture which supports this interpretation: There exists
an extremal, radially outgoing null geodesic $\gamma^\ast$
in the outer domain $D$ such that $\lim_{s\to \infty}\gamma^\ast (s) = 2 M$
where $M$ is the ADM mass of the solution, and  as $t\to\infty$
the metric converges for $r>2 M$ to the Schwarzschild metric 
representing a black hole of mass $M$. 

In the next section we state our main results for
the Einstein-Vlasov system, where we specify classes
of spherically symmetric initial data which lead to solutions
showing the above behaviour. The Vlasov equation and the
corresponding energy-momentum tensor components
in the case of spherical symmetry are stated there.
In Section \ref{secgenmat} we give a general formulation of one of our results
where no particular matter model is considered. The reason for this
is that most steps in the proof of Theorem~\ref{vlasov2} below are
of a general character and---besides the fact that for the
Einstein-Vlasov system there is an existence theory for the initial
value problem which guarantees the existence of solutions on $D$---the
specific properties of Vlasov matter are used only in one key lemma.
Hence it is natural to precisely single out the required conditions on
the level of the macroscopic matter quantities. This clarifies
the main mechanism in our method, and it may lead to applications of
our method to other matter models.
Using an additional feature of Vlasov matter we construct
an alternative, and in some respects larger, class of initial
data which ensure
the formation of black holes, cf.\ Theorem~\ref{vlasov1}.

The proofs of our results then proceed as follows. After stating some
general auxiliary results in Section~\ref{prelim} we prove
Theorem \ref{genmat}, which is the general-matter version
of Theorem \ref{vlasov2}, in Section~\ref{secgenmatproof}.
The latter result is then established in Section~\ref{secvlasov2}
by showing that Vlasov matter satisfies the required general conditions
on the matter for a suitable class of initial data. Theorem~\ref{vlasov1}
is established in Section~\ref{secvlasov1} together with Corallary~\ref{ssinthemiddle} 
on global existence in Schwarzschild coordinates. 
For all these results it is essential to make sure that in the
outer region $D$ all the matter moves inward. In the case of general matter
this is a condition which we have to impose on the solution,
whereas in the case of Vlasov matter we can specify conditions on the initial
data such that this is true. In Section~\ref{bhproof} we prove the
convergence of our solution to a Schwarzschild black hole of the
corresponding ADM mass in the case of Vlasov matter.
%
%
\section{Main results for Vlasov matter} \label{secvlasres}
\setcounter{equation}{0}
In this section Eqns.\ (\ref{boundc})--(\ref{ein3}) will be supplemented by
the spherically symmetric version of the Vlasov equation together with expressions
for the relevant components of the energy-momentum tensor so that a
closed system is obtained, known as the spherically symmetric, asymptotically
flat Einstein-Vlasov system.

In order to exploit the symmetry it is useful to introduce
non-canonical variables on momentum space and write $f=f(t,r,w,L)$.
For a detailed derivation of the corresponding
equations we refer to \cite{Rein95}; here we just state the result.
The Vlasov equation is
\begin{equation} \label{vlasov}
\partial_{t}f+e^{\mu-\lambda}\frac{w}{E}\partial_{r}f
-\left(\lambda_{t}w+e^{\mu-\lambda}\mu_{r}E-
e^{\mu-\lambda}\frac{L}{r^3E}\right)\,\partial_{w}f=0,
\end{equation}
where
\[
E=E(r,w,L):=\sqrt{1+w^{2}+L/r^{2}} = e^\mu p^0.
\]
The variables $w\in ]-\infty,\infty[$ and $L\in [0,\infty[$ can be
thought of as the radial component of the momentum and the square of
the angular momentum respectively.
Notice that the latter
is conserved along characteristics of the Vlasov equation.
The matter quantities are given by
\begin{eqnarray}
\rho(t,r)
&=&
e^{-2 \mu} T_{00}(t,r) =
\frac{\pi}{r^{2}}
\int_{-\infty}^{\infty}\int_{0}^{\infty}Ef(t,r,w,L)\,dL\,dw,
\label{rho}\\
p(t,r)
&=&
e^{-2 \lambda} T_{11}(t,r) =
\frac{\pi}{r^{2}}\int_{-\infty}^{\infty}\int_{0}^{\infty}
\frac{w^{2}}{E}f(t,r,w,L)\,dL\,dw,
\label{p}\\
j(t,r)
&=&
-e^{-(\lambda+\mu)} T_{01}(t,r) =
\frac{\pi}{r^{2}}
\int_{-\infty}^{\infty}\int_{0}^{\infty}w\,f(t,r,w,L)\,dL\,dw.
\label{j}
\end{eqnarray}
Notice that the quantities $\rho,\, p,\, j$ appear on the right hand
sides of the field equations (\ref{ein1})--(\ref{ein3}),
and they are given in terms of $f$ alone, which is the main reason
for using the non-canonical variables $w$ and $L$.
The system (\ref{boundc})--(\ref{ein3}), (\ref{vlasov})--(\ref{j}) is
the spherically symmetric Einstein-Vlasov system in Schwarzschild coordinates.
As initial data we need to prescribe an initial distribution function
$\open{f}=\open{f}(r,w,L)\geq 0$, which should be compactly supported in
$]0,\infty[ \times ]-\infty,\infty[\times ]0,\infty[$, and such that
\begin{equation}\label{notsinit}
   \int_0^r 4\pi\eta^2\open{\rho}(\eta)\,d\eta
   =4\pi^2 \int_0^r\int_{-\infty}^{\infty}\int_0^{\infty}
   E\open{f}(\eta,w,L)\,dL\,dw\,d\eta < \frac{r}{2}.
\end{equation}
The origin $r=0$ is excluded from the support for technical
reasons, but this could be avoided by using Cartesian coordinates.
The condition (\ref{notsinit}) implies that the equations (\ref{ein1}) and
(\ref{ein2}) have solutions $\lambda$ and $\mu$, cf.~Section~\ref{prelim},
and since $\open{f}$ has compact support, a property which is inherited by
$f(t)$, the matter terms are well defined.
If in addition $\open{f}$ is  $C^1$ we say that the initial data is
\textit{regular}. As is shown in \cite{RR1} or \cite{Rein95},
regular initial data launch a unique local solution for which
all the derivatives which appear in the system exist classically.
In Section~\ref{secvlasov2} we discuss in more detail that
this local solution extends to the whole outer region $D$
defined in (\ref{ddef}).

To state our main results
let $0<r_0<r_1$ be given, put $M=r_1/2$ (this is going to be the ADM
mass of the solution), and fix $0<M_\mathrm{out}<M$ such that
\begin{equation}\label{icnts}
   \frac{2(M-M_\mathrm{out})}{r_0}<\frac{8}{9}.
\end{equation}
\textbf{Remark.} The value $8/9$ is chosen for definiteness,
and any number less than one would do, effecting the values of some
of the constants below.
\smallskip

Two different theorems will be stated below, corresponding
to the following two situations.
\begin{itemize}
\item[(i)]
Let $R_1>r_1$ be such that
\begin{equation}\label{mediumstrip}
   R_1-r_1<\frac{r_1-r_0}{6},
\end{equation}
or
\item[(ii)]
let $R_1>r_1$ be such that
\begin{equation}\label{smallstrip}
   \sqrt{\frac{R_1-r_1}{R_1}}<\min\left\{\frac{1}{6},\frac{r_0^2}{12\kappa R_1M},
   \frac{r_1-r_0}{8\kappa R_1}\right\},
\end{equation}
where the (explicit) constant $\kappa>0$ will be specified
in Theorems \ref{vlasov2} and \ref{genmat} below.
\end{itemize}
Finally, we define
\[
R_0:=\frac{1}{2}(r_1+R_1).
\]
Denote by $\open{\rho}\,$ the energy density induced by the initial
distribution function $\open{f}$. We require that all the matter
in the outer region $[r_0, \infty[$ is initially located in the strip $[R_0,R_1]$,
with $M_\mathrm{out}$ being the corresponding fraction of the ADM mass $M$, i.e.,
\begin{equation}\label{checkM}
\int_{r_0}^{\infty}4\pi r^2\open{\rho}(r)dr = \int_{R_0}^{R_1}4\pi r^2\open{\rho}(r)dr=M_\mathrm{out}.
\end{equation}
Furthermore, the remaining fraction $M-M_\mathrm{out}$ should be
initially located within the ball of area radius $r_0$, i.e.,
\begin{equation}\label{M-checkM}
\int_{0}^{r_0}4\pi r^2\open{\rho}(r)dr=M-M_\mathrm{out}.
\end{equation}
\textbf{Remark.} The set up described above is quite similar
to the set up in \cite{Chr91} for a scalar field.
In \cite{Chr91} it is not required to have matter in an
``inner'' strip $[0,r_0]$, as is the case here in view of (\ref{M-checkM})
and the condition $M_\mathrm{out}<M.$ The reason why we need some matter in
the region $r\le r_0$ is to ensure that initially ingoing matter continues
to be ingoing for all times, cf.~Lemma \ref{ingoinglemma} below.
If one only considers purely radially ingoing particles,
i.e., with no angular momentum (which results in a non-smooth distribution function $f$),
then we could allow for $M_\mathrm{out}=M.$ It is interesting to note that $p=\rho$ holds
for Vlasov matter, if the particles have no angular momentum and their rest mass is zero,
which is the case for the scalar field considered in \cite{Chr91}.
\smallskip

Now we are in the position to formulate our main results for Vlasov matter.
Corresponding to Case (i) above, we prove
\begin{theorem} \label{vlasov1}
Let $r_0,\, r_1, M$, and $M_\mathrm{out}$ be given as above, and let $R_1$
satisfy (\ref{mediumstrip}).
Then there exists a  set ${\cal{I}}_1$ of regular initial data for
the spherically symmetric Einstein-Vlasov system such that if
$\open{f}\in{\cal{I}}_1$, then (\ref{checkM}) and (\ref{M-checkM}) hold,
the corresponding solution exists on $D$,
and
\[
\lim_{s\to \infty}\gamma^+(s) < \infty,
\quad \lim_{s\to \infty} \int_{\gamma^+(s)}^\infty 4 \pi r^2 \rho(s,r)\,dr > 0,
\]
where $\gamma^+$ satisfies (\ref{gamma+}).
\end{theorem}
By abuse of notation we denote by $D$ both the outer region
in spacetime defined by (\ref{ddef}) and the part of the
mass shell with $(t,r) \in D$.

The next theorem addresses Case (ii) above,
assuming the stronger condition (\ref{smallstrip}).
This allows for a more straightforward proof,
and the constraints on the momentum variables
of the initial distribution function $\open{f}$ which are used
to specify the set ${\cal{I}}_1$ will be slightly relaxed.
Hence, the initial data set ${\cal{I}}_1$ does not contain ${\cal{I}}_2$ in
Theorem \ref{vlasov2} below,
but it is larger in the sense that data in ${\cal{I}}_2$ are
quite close to containing a trapped surface, which is not necessarily
the case for data in ${\cal{I}}_1$.
The precise form of ${\cal{I}}_1$ and ${\cal{I}}_2$ is specified in the proofs.

\begin{theorem} \label{vlasov2}
Let $r_0,\, r_1, M$, and $M_\mathrm{out}$ be given as above and let $R_1$ satisfy
(\ref{smallstrip}) with $\kappa=6$.
Then there exists a  set ${\cal{I}}_2$ of regular initial data for
the spherically symmetric Einstein-Vlasov system such that if
$\open{f}\in{\cal{I}}_2$, then (\ref{checkM}) and (\ref{M-checkM})
hold, the corresponding solution exists on $D$, and
\[
\lim_{s\to \infty}\gamma^+(s) < \infty,
\quad \lim_{s\to \infty} \int_{\gamma^+(s)}^\infty 4 \pi r^2 \rho(s,r)\,dr > 0,
\]
where $\gamma^+$ satisfies (\ref{gamma+}).
\end{theorem}
The Einstein-Vlasov system has a wide variety of static, spherically
symmetric  solutions with
finite ADM mass and finite radius, i.e., compact support of the matter,
cf.\ \cite{Rss1, RRss1, RRss2}. Particularly interesting examples of initial
data for which our results apply are obtained if the matter for $r \leq r_0$
is represented by such a static solution, more precisely:

\begin{cor} \label{ssinthemiddle}
Let $f_s$ be a static solution of the spherically symmetric Einstein-Vlasov
system with finite ADM mass $M_s>0$ and finite radius $r_s >0$. Define
$r_0=r_s$, let $r_1 > r_0$ be arbitrary, $M=r_1/2$, and 
$M_\mathrm{out} = M -M_s$; the latter quantity is positive. 
Then the initial data sets ${\cal{I}}_1$ and
${\cal{I}}_2$ both contain data $\open{f}$ which coincide with the
given static solution for $0\leq r\leq r_0$. The corresponding solution $f$
of the Einstein-Vlasov system exists for all $r\geq 0,\ t\geq 0$
and coincides with the static solution $f_s$
for all $r\leq \gamma^+(t)$ and $t\geq 0$.
\end{cor}

We prove this result at the end of Section~\ref{secvlasov1}. 
It represents a global existence result for the Einstein-Vlasov
system in Schwarzschild time for data which are not small.

In the next section we formulate a version of Theorem~\ref{vlasov2} for quite general
matter models. One reason for this is that the main mechanism behind our method
becomes very transparent by posing sufficient conditions on the macroscopic
matter terms rather than conditions on the initial distribution function $\open{f}$ as
we did in the theorems above. Theorem~\ref{vlasov2} will then be a consequence
of this generalization, cf.~Section~\ref{secvlasov2}, whereas
Theorem \ref{vlasov1} is established in Section \ref{secvlasov1}.

In these proofs it turns out that for the classes of initial data
that we specify we can obtain somewhat sharper asymptotic information
on $\gamma^+$ and the mass in the outer region; see (\ref{precise}) below.
More importantly, we can establish the following
additional information which shows that the solution evolves towards
a Schwarzschild black hole of mass $M$. 

\begin{theorem} \label{bh}
In the situation of Theorem~\ref{vlasov1} or Theorem~\ref{vlasov2} the following holds:
\begin{itemize}
\item[(a)]
There exist constants $\alpha, \beta > 0$ depending only on the
initial data set ${\cal I}_1$ or ${\cal I}_2$ respectively such
that if
\[
t \geq 0 \ \mbox{and}\ r \geq 2 M + \alpha e^{-\beta t}
\]
then $f(t,r,\cdot,\cdot) = 0$, i.e., we have vacuum, and the metric equals
the Schwarzschild metric 
\[
ds^2 = -\left(1-\frac{2 M}{r}\right)\, dt^2 + \left(1-\frac{2 M}{r}\right)^{-1}dr^2\,
+ r^2(d\theta^2 + \sin^2 \theta d\varphi^2),
\]
representing a black hole of mass $M$. 
\item[(b)]
For all $t\geq 0$ and $\gamma^+ (t) \leq r \leq 2 M + \alpha e^{-\beta t}$,
\[
\mu(t,r) 
\leq \ln \left(\frac{\alpha e^{-\beta t}}{2 M + \alpha e^{-\beta t}}\right)^{1/2} 
\]
so that in the outer region $D$,
\[
\lim_{t\to \infty} \mu(t,r) = -\infty\ \mbox{for}\
r \leq 2 M,
\] 
and the timelike lines $r=c,$ where $c\in [0,2M],$ are incomplete and their 
proper lengths are uniformly bounded by a constant depending on $\alpha,\, \beta$ and $M.$
 \item[(c)]
Let
\beas
r^\ast := \sup \{ r\geq r_0
&|&
\mbox{the radially outgoing null geodesic}\ \gamma\ \mbox{with}\\
&&
\gamma(0)=r\ \mbox{satisfies}\ 
\lim_{s\to\infty}\gamma (s) < \infty \},
\eeas
and let $\gamma^\ast$ be the radially outgoing null geodesic
with $\gamma^\ast (0) = r^\ast$. Then
\[
\lim_{s\to\infty}\gamma^\ast (s) = 2 M,
\]
and every  radially outgoing null geodesic $\gamma$ with $\gamma(0) > r^\ast$
is future complete with $\lim_{s\to\infty}\gamma (s) = \infty$.
\end{itemize}
\end{theorem}

\section{The result for general matter models} \label{secgenmat}
\setcounter{equation}{0}
In this section we specify the general assumptions on a matter model
sufficient for our method to be applied. In order to keep the discussion
consistent with the Vlasov part of our arguments,
and in view of the right hand sides of the field equations
(\ref{ein1}), (\ref{ein2}), (\ref{ein3}), it is convenient to use the notation
\begin{equation} \label{genmatquant}
   \rho := e^{-2\mu} T_{00},\ p := e^{-2\lambda} T_{11},\
   j := - e^{-\mu-\lambda}T_{01}.
\end{equation}
Firstly, we  assume that the following two conditions
are satisfied.
\begin{itemize}
\item The dominant energy condition holds. \hfill(DEC)
\item The radial pressure $p$ is non-negative. \hfill(NNP)
\end{itemize}
The dominant energy condition (DEC) plays a central role in general
relativity and is the main criterion that a matter model should satisfy
to be considered realistic. We refer to \cite{HE} for its definition.
The non-negative pressure condition (NNP) is restrictive in the sense
that it rules out, for example, a Maxwell field as matter model.
However, for most astrophysical models it is a standard assumption,
with e.g.~fluid models satisfying this condition.
For the purpose of this paper we only need to focus
on two consequences of these two criteria, cf.\ \cite{HE} and \cite{P}.
The (DEC) condition implies, together with the (NNP) condition, that
\begin{equation}\label{nnrho}
   0\leq p\leq \rho\ \mbox{and}\ |j| \leq \rho.
\end{equation}
Furthermore, by (DEC) any geodesic $(s,R(s))$ of a material particle
or a light ray satisfies
\begin{equation}\label{mattergeodesic}
   \left|\frac{dR(s)}{ds}\right| \leq e^{(\mu-\lambda)(s,R(s))}.
\end{equation}
The meaning of the latter condition is that locally the speed of energy flow is
less than or equal to the speed of light.

Let $\lambda,\, \mu,\, \rho,\, p,\, j$ correspond to a solution
of the spherically symmetric Einstein-matter equations
(\ref{boundc})--(\ref{ein3}), (\ref{matevol}), (\ref{emtdef})
in Schwarzschild coordinates, launched by initial data from
a class  $\cal{I}$.
In order to investigate the global structure of the solutions it is
necessary that they exist globally in an appropriate sense.
In the situation at hand they need to exist on the outer region
$D$ defined in (\ref{ddef}).
In the spherical symmetric case the main obstruction for obtaining global
solutions arises from the difficulties related to the centre of
symmetry $r=0$. For example, for a massless scalar field or a collisionless
gas as matter model it has been shown that solutions remain regular
away from $r=0$ for general initial data, cf.\ \cite{Chr99,And1,RRS}.
On the other hand, for dust a singularity of shell crossing type
can also occur at some $r>0$. Although in that case there are no true
geometric spacetime singularities, such behaviour has to be ruled out
in order not to interfere with the analysis of the solution on $D$.
This can be achieved by proper assumptions on the initial data, cf.~\cite{Chr84}.
In view of (\ref{mattergeodesic}) a possible break down of
solutions at $r=0$ will have no influence on the outer domain $D$.
Hence we formulate a third condition, concerning global existence of
solutions in the outer domain, as follows.
\begin{itemize}
\item
For solutions launched by data from the set $\cal{I}$,
$\gamma^+$ defined by (\ref{gamma+}) exists on $[0,\infty[$, and
$\lambda,\, \mu,\, \rho,\, p,\, j\in C^1(D).$ \hfill(GLO)
\end{itemize}
The three conditions above are of a quite general nature.
The fourth and final condition however,
is  tightly connected to our method of proof.
\begin{itemize}
\item There exists a constant $c_1>0$ such that
$\rho \leq -c_1 j$ in $D$. \hfill(GCC)
\end{itemize}
The acronym (GCC) stands for ``gravitational collapse condition'',
and this condition plays a crucial role for our method of proof.
We emphasize that our main results show that for Vlasov matter
there are initial data sets such that (GCC) holds.
As a first consequence of (GCC) and (\ref{nnrho}),
note that $j\le 0$ in $D$, i.e., the matter is ingoing for all times.
In this respect our present results complement \cite{AKR1},
where purely outgoing matter is considered.
\smallskip

Let us now assume that our matter model satisfies (DEC) and (NNP),
and that there exists an initial data set $\cal{I}$ such that (GLO) and
(GCC) hold as well. Then we have the following result, which should be viewed
as a version of Theorem \ref{vlasov2} for general matter.
\begin{theorem} \label{genmat}
Let $r_0,\, r_1,\, M$, and $M_\mathrm{out}$ be given as above
and let $R_1$ satisfy (\ref{smallstrip}) with $\kappa=2c_1$.
Assume that there exists an initial data set
${\cal{I}}_3\subset\cal{I}$ such
that (\ref{checkM}) and (\ref{M-checkM})
hold for all initial data in ${\cal{I}}_3$. Then
for any solution launched by initial data in ${\cal{I}}_3$,
\[
\lim_{s\to \infty}\gamma^+(s) < \infty,
\quad \lim_{s\to \infty} \int_{\gamma^+(s)}^\infty 4 \pi r^2 \rho(s,r)\,dr > 0,
\]
where $\gamma^+$ satisfies (\ref{gamma+}).
\end{theorem}
The detailed information on the gravitational collapse
which for Vlasov matter is provided in Theorem~\ref{bh}
is not available in the present situation,
but the following still holds.

\smallskip

\noindent
{\bf Remark}. In the situation of Theorem~\ref{genmat},
\[
\lim_{t\to \infty} \mu(t,r) = -\infty\ \mbox{for}\
\lim_{s\to\infty}\gamma^+(s) \leq r \leq r_1
\]
for some $r_1 > \lim_{s\to\infty}\gamma^+(s)$. If
$r^\ast$ and $\gamma^\ast$ are defined as in Theorem~\ref{bh}
then
\[
\lim_{s\to\infty}\gamma^\ast (s) < \infty,
\]
and every  radially outgoing null geodesic $\gamma$ with $\gamma(0) > r^\ast$
is future complete with $\lim_{s\to\infty}\gamma (s) = \infty$.

These assertions will be established in Section~\ref{bhproof}.
Concerning the question which matter models besides Vlasov matter
satisfy our conditions above we note the following:

\noindent
{\bf Remark}. For a spherically symmetric perfect fluid with density
${\cal R}$, pressure $P=P({\cal R})$, and radial velocity field $u$,
the (DEC) and (NNP) conditions and Eqn.~(\ref{nnrho}) respectively
are satisfied provided that $0\leq P({\cal R}) \leq  {\cal R}$,
which restricts the equation of state. The (GCC) condition holds
for example with $c_1=\sqrt{2}$ if $-e^{\lambda} u \geq 1$ on $D$.
In the kinetic context of the Vlasov model we derive analogous
estimates on the particle level from conditions on the initial data.
\section{Preliminaries} \label{prelim}
\setcounter{equation}{0}
In this section we collect some general facts concerning
the spherically symmetric Einstein-matter equations
under the assumptions (DEC) and (NNP) that have been specified
in the previous section.

A quantity which plays an important role is the quasi-local mass $m(t,r)$.
Typically, the spherically symmetric Einstein-matter system is
supplemented by the
requirement of a regular centre, i.e., $\lambda (t,0)=0$.
Using this boundary condition the field equation (\ref{ein1}) implies that
\begin{equation}\label{e2lamb}
   e^{-2\lambda}=1-\frac{2m}{r},
\end{equation}
where the quasi-local mass would be given by
$m(t,r) := \int_0^r 4\pi\eta^2\rho(t,\eta)\,d\eta$.
Then $m(t,\infty)$ is a conserved quantity, the ADM mass.
However, in the present context we want to investigate
the system on the outer domain $D$, regardless of whether or not
the solution remains regular in the region $r<\gamma^+(t)$.
Hence we do not use the usual boundary condition at $r=0$.
Instead, we assume that the ADM mass $M>0$ is given and redefine
the quasi-local mass by
\begin{equation}\label{m-def}
   m(t,r)=M-\int_r^\infty 4\pi\eta^2\rho(t,\eta)\,d\eta.
\end{equation}
Then $\lim_{r\to\infty} m(t, r)=M$, $0\le m\le M$, and $m_r=4\pi r^2\rho$
holds. Defining $\lambda$ by (\ref{e2lamb}), (\ref{genmatquant}) shows that
(\ref{ein1}) and the boundary condition in (\ref{boundc}) are satisfied.
In addition, we need to modify (\ref{notsinit}) to
\begin{equation}\label{notsinit2}
   \open{m}(r)<\frac{r}{2},\quad r\in ]0, \infty[,
\end{equation}
a condition that once again will be included in the notion
of regular initial data.

By (\ref{ein1}) and (\ref{ein2}),
\begin{equation} \label{mur}
\lambda_r=\Big(4\pi r\rho-\frac{m}{r^2}\Big)e^{2\lambda},
   \quad\mu_r=\Big(\frac{m}{r^2}+4\pi r p\Big)e^{2\lambda}.
\end{equation}
In view of (\ref{boundc}), $\mu=\hat{\mu}+\check{\mu}$, where we define
\begin{eqnarray}
\hat{\mu}(t, r)
&:=&
-\int_r^\infty\frac{m(t, \eta)}{\eta^2}\,e^{2\lambda(t,\,\eta)}\,d\eta,\label{hatmu}
\\
\check{\mu}(t, r)
&:=&
-\int_r^\infty 4\pi\eta\,p(t, \eta)\,e^{2\lambda(t,\,\eta)}\,d\eta.
\label{checkmu}
\end{eqnarray}
\begin{lemma}\label{hatmu-lem}
The following assertions hold.
\begin{itemize}
\item[(a)]
$2\hat{\mu}\le\mu-\lambda\le\hat{\mu}\le\hat{\mu}+\lambda$.
\item[(b)]
$\mu+\lambda\le\hat{\mu}+\lambda$.
\item[(c)]
$(\mu-\lambda)(t, r)=2\hat{\mu}(t, r)
+\int_r^\infty 4\pi\eta\,(\rho-p)(t, \eta)\,e^{2\lambda(t,\,\eta)}\,d\eta$.
\item[(d)]
$\hat{\mu}_t(t, r)=\int_r^\infty 4\pi j(t, \eta)
\,e^{(\mu+\lambda)(t,\,\eta)}e^{2\lambda(t,\,\eta)}\,d\eta$.
In particular, if $j\le 0$, then also $\hat{\mu}_t\le 0$.
\end{itemize}
\end{lemma}
{\bf Proof\,:} In view of (\ref{boundc}),
\[
\lambda(t, r)=-\int_r^\infty\Big(4\pi\eta\,\rho(t, \eta)
   -\frac{m(t, \eta)}{\eta^2}\Big)\,e^{2\lambda}\,d\eta
= - \int_r^\infty 4\pi\eta\,\rho(t, \eta)\,e^{2\lambda}\,d\eta -\hat{\mu} ,
\]
and by (\ref{nnrho}) the relation $\mu-\lambda\ge 2\hat{\mu}$ follows.
On the other hand,
by (\ref{e2lamb}), $\lambda\ge 0$. Thus $\check{\mu}\le 0$
leads to $\mu-\lambda\le\mu\le\hat{\mu}\le\hat{\mu}+\lambda$, and part (a)
is established.
Part (b) follows from $\check{\mu}\le 0$. As to (c), we observe that
\[
\hat{\mu}+\lambda+\int_r^\infty
4\pi\eta\,(\rho-p)\,e^{2\lambda}\,d\eta=\check{\mu},
\]
which gives the claim. By (\ref{e2lamb}) and (\ref{ein3}),
${(e^{2\lambda}\frac{m}{r^2})}_t
=\frac{1}{2r}{(e^{2\lambda}-1)}_t=-4\pi\,e^{\mu+\lambda}e^{2\lambda}j$. Hence (d)
follows from (\ref{hatmu}).
\prfe
\begin{lemma}\label{int0infty} For $r\in [0, \infty[$ the following holds:
\begin{eqnarray*}
   & & \int_r^\infty 4\pi\eta\,(\rho+p)(t, \eta)
   \,e^{(\mu+\lambda)(t,\,\eta)}e^{2\lambda(t,\,\eta)}\,d\eta
   =1-e^{(\mu+\lambda)(t,\,r)}\le 1,
   \\ & & \int_r^\infty 4\pi\eta\,\rho(t, \eta)
   \,e^{(\hat{\mu}+\lambda)(t,\,\eta)}e^{2\lambda(t,\,\eta)}\,d\eta
   =1-e^{(\hat{\mu}+\lambda)(t,\,r)}\le 1.
\end{eqnarray*}
\end{lemma}
{\bf Proof\,:} It suffices to integrate
\begin{eqnarray}\label{gsto}
\partial_r(e^{\mu+\lambda})
& = &
e^{\mu+\lambda}(\mu_r+\lambda_r) = e^{\mu+\lambda} 4\pi r\,(p+\rho),
\nonumber \\
\partial_r(e^{\hat{\mu}+\lambda})
& = &
e^{\hat{\mu}+\lambda}(\hat{\mu}_r+\lambda_r)
=e^{\hat{\mu}+\lambda}\Big(e^{2\lambda}\frac{m}{r^2}
+\Big(4\pi r\rho-\frac{m}{r^2}\Big)e^{2\lambda}\Big)
\nonumber \\
& = &
4\pi r\rho\,e^{\hat{\mu}+\lambda}e^{2\lambda},
\end{eqnarray}
observing that $\lim_{r\to\infty}\hat{\mu}(t, r)
=\lim_{r\to\infty}\lambda(t, r)=\lim_{r\to\infty}\mu(t, r)=0$.
For Vlasov matter, the first relation has been used in \cite[Lemma 1]{And1}.
\prfe
Next we consider outgoing and ingoing radial null geodesics
$\gamma^+$ and $\gamma^-$, respectively.
\begin{lemma}\label{gammapm}
Let $\gamma^{\pm}$ be the solutions to
\[
\frac{d\gamma^{\pm}}{ds}(s)=\pm\,e^{(\mu-\lambda)(s,\,\gamma^{\pm}(s))},
   \quad\gamma^+(0)=r_0<r_1=\gamma^-(0).
\]
Then
\begin{itemize}
\item[(a)]
$\gamma^+$ is strictly increasing, $s\mapsto m(s, \gamma^+(s))$ is increasing,
and the limits $\lim_{s\to\infty}\gamma^+(s)\in ]r_0, \infty]$
and $\lim_{s\to\infty} m(s, \gamma^+(s))\in [m(0, r_0), M]$ exist.
\item[(b)]
$\gamma^-$ is strictly decreasing, $s\mapsto m(s, \gamma^-(s))$ is decreasing,
and the limits $\lim_{s\to\infty}\gamma^-(s)\in [0, r_1[$
and $\lim_{s\to\infty} m(s, \gamma^-(s))\in [0, m(0, r_1)]$ exist.
\item[(c)]
The relation
\[
\frac{d}{ds}(\hat{\mu}+\lambda)(s, \gamma^{\pm}(s))
   =\Big(\hat{\mu}_t-4\pi r\,e^{\mu+\lambda}(j\mp\rho)\Big)
   \bigg|_{(t,\,r)=(s, \gamma^{\pm}(s))}
\]
holds. In particular,
if $j\le 0$ and $\rho=j=0$ along $\gamma^{\pm}$,
then also $\frac{d}{ds}(\hat{\mu}+\lambda)(s, \gamma^{\pm}(s))\le 0$.
\end{itemize}
\end{lemma}
{\bf Proof\,:} Differentiating (\ref{e2lamb}) w.r.t.\ $t$
and using (\ref{ein3}) implies that $m_t=-4\pi r^2 e^{\mu-\lambda} j$.
Since $\rho\ge j$ according to (\ref{nnrho}), this
yields
\begin{eqnarray*}
   \frac{d}{ds}\,m(s, \gamma^+(s)) & = & m_t(s, \gamma^+(s))
   +m_r(s, \gamma^+(s))\frac{d\gamma^+}{ds}(s)
   \\ & = & (-4\pi r^2 e^{\mu-\lambda} j
   +4\pi r^2\rho\,e^{\mu-\lambda})\big|_{(t,\,r)=(s, \gamma^+(s))}\ge 0.
\end{eqnarray*}
Thus part (a) is obtained from $m\le M$.
Since $\rho\ge -j$, the proof of (b) is analogous to (a).
As to (c), note that by definition of $\hat{\mu}$, (\ref{ein1}), and (\ref{ein3}),
\begin{eqnarray*}
   \lefteqn{\frac{d}{ds}(\hat{\mu}+\lambda)(s, \gamma^{\pm}(s))}
   \\ & = & \Big(\hat{\mu}_t+\hat{\mu}_r\frac{d\gamma^{\pm}}{ds}
   +\lambda_t+\lambda_r\frac{d\gamma^{\pm}}{ds}\Big)\bigg|_{(t,\,r)=(s, \gamma^{\pm}(s))}
   \\ & = & \Big(\hat{\mu}_t\pm\frac{m}{r^2}\,e^{2\lambda}e^{\mu-\lambda}
   -4\pi r\,e^{\mu+\lambda}j\pm\Big(4\pi r\rho-\frac{m}{r^2}\Big)e^{2\lambda}
   e^{\mu-\lambda}\Big)\bigg|_{(t,\,r)=(s, \gamma^{\pm}(s))}
   \\ & = & \Big(\hat{\mu}_t-4\pi r\,e^{\mu+\lambda}(j\mp\rho)\Big)
   \bigg|_{(t,\,r)=(s, \gamma^{\pm}(s))},
\end{eqnarray*}
as desired. The last claim follows from Lemma~\ref{hatmu-lem}(d).
\prfe
%
%
\section{Proof of Theorem~\ref{genmat}} \label{secgenmatproof}
\setcounter{equation}{0}
%
%
In this section we use the hypotheses stated in Section~\ref{secgenmat} to
prove Theorem~\ref{genmat}.
The proof is short and emphasizes that the crucial mechanism is captured in
the (GCC) condition. Our main results which  show in particular that the (GCC)
condition holds for Vlasov matter are established in the next sections.

Consider the out- and ingoing
null geodesics $\gamma^+$ and $\gamma^-$
defined in Lemma~\ref{gammapm}. The claims follow if we can show
that these geodesics never intersect.
By continuity and monotonicity there exists $T \in ]0,\infty]$ such that
\begin{equation}\label{fg-esti}
   r_0\le\gamma^+(t) < \gamma^-(t)\le r_1,\quad t\in [0, T[;
\end{equation}
it will be shown that actually $T=\infty$ holds.
In view of (\ref{checkM}) we have initially that
$\rho=p=j=0$ for $r\geq R_1.$ The (GCC) condition implies that $j\leq 0$
in $D$, meaning that the flow of matter is ingoing. Therefore
\begin{equation}\label{ingoing}
   \rho=p=j=0\quad\mbox{and}\quad m=M
   \quad\mbox{for}\quad (t,r)\in [0, T[\times [R_1, \infty[.
\end{equation}
By Lemma \ref{int0infty}, (\ref{nnrho}),
the (GCC) condition, and Lemma~\ref{hatmu-lem}(d) for $s\in [0,T[$
and $r\in [\gamma^+(s), \infty[$,
\begin{eqnarray*}
   1-e^{(\mu+\lambda)(s,\,r)} & = & \int_r^\infty 4\pi\eta\,(\rho+p)(s, \eta)
   \,e^{(\mu+\lambda)(s,\,\eta)}e^{2\lambda(s,\,\eta)}\,d\eta
   \\ & \le & 2c_1\int_r^\infty 4\pi\eta\,|j(s, \eta)|
   \,e^{(\mu+\lambda)(s,\,\eta)}e^{2\lambda(s,\,\eta)}\,d\eta
   \\ & \le & -2c_1 R_1\int_r^\infty 4\pi j(s, \eta)
   \,e^{(\mu+\lambda)(s,\,\eta)}e^{2\lambda(s,\,\eta)}\,d\eta
   \\ & = & -2c_1 R_1\hat{\mu}_t(s, r),
\end{eqnarray*}
since $j(s, \eta)\neq 0$ implies $\eta\le R_1$. Thus
\begin{equation}\label{mut}
   \hat{\mu}_t(s, r)\le
   -\frac{1}{2c_1 R_1}\Big(1-e^{(\mu+\lambda)(s,\,r)}\Big).
\end{equation}
This in turn implies that
\begin{eqnarray}\label{est}
   \lefteqn{\hat{\mu}(t, \gamma^{\pm}(t))-\hat{\mu}(0, \gamma^{\pm}(0))}\nonumber
   \\ & = & \int_{0}^t\frac{d}{ds}\,\hat{\mu}(s, \gamma^{\pm}(s))\,ds\nonumber
   \\ & = & \int_{0}^t\Big(\hat{\mu}_t(s, \gamma^{\pm}(s))
   \pm\hat{\mu}_r(s, \gamma^{\pm}(s))e^{(\mu-\lambda)(s,\,\gamma^{\pm}(s))}\Big)\,ds\nonumber
   \\ & \le & \int_{0}^t\Big(-\frac{1}{2c_1 R_1}\Big(1-e^{(\mu+\lambda)(s,\,\gamma^{\pm}(s))}\Big)
   \pm\frac{m(s,
   \gamma^{\pm}(s))}{\gamma^{\pm}(s)^2}\,e^{(\mu+\lambda)(s,\,\gamma^{\pm}(s))}\Big)\,ds
   \nonumber
   \\ & \le & -\frac{t}{2c_1 R_1}
   +\int_{0}^t\Big(\frac{1}{2c_1 R_1}
   +\frac{m(s, \gamma^{\pm}(s))}{\gamma^{\pm}(s)^2}\Big)\,e^{(\mu+\lambda)(s,\,\gamma^{\pm}(s))}\,ds.
\end{eqnarray}
Now for any $r\in [r_0,r_1]$ and $t\in [0,T[$ it follows from $\hat{\mu}_r\ge 0$
and (\ref{e2lamb}) that
\begin{equation}
   \hat{\mu}(t,r)\leq \hat{\mu}(t,R_1)
   =-\int_{R_1}^{\infty}\frac{M\,d\eta}{\eta^2(1-2M/\eta)}.
\end{equation}
Using $M=r_1/2$ we get
\[
\hat{\mu}(t,R_1)=\frac{1}{2}\log\Big(\frac{R_1-r_1}{R_1}\Big),
\]
so that for $r\in [r_0,r_1],$
\begin{equation}\label{emuhat}
   e^{\hat{\mu}(t,\,r)}\leq e^{\hat{\mu}(t,R_1)}=\sqrt{\frac{R_1-r_1}{R_1}}.
\end{equation}
By (\ref{mattergeodesic}) and the properties of the initial matter distribution
there is vacuum in the region $\gamma^+(t) \leq r\leq\gamma^-(t)$.
Hence $m(t, r)=M-M_\mathrm{out}$ and (\ref{icnts}) imply that
\begin{equation} \label{explambdaless3}
e^{\lambda(t,r)}\leq\frac{1}{\sqrt{1-2(M-M_\mathrm{out})/r_0}}<3
\end{equation}
for $\gamma^+(t) \leq r\leq\gamma^-(t)$. From Lemma \ref{hatmu-lem}(b)
and (\ref{smallstrip}), recalling $\kappa=2c_1$,
we obtain in particular that
\[
e^{(\mu+\lambda)(s,\gamma^{\pm}(s))}\leq
e^{(\hat{\mu}+\lambda)(s,\gamma^{\pm}(s))} <
\min\left\{\frac12,\frac{r_0^2}{8c_1R_1M}\right\} =: d.
\]
Thus (\ref{est}) yields
\begin{eqnarray*}
   \hat{\mu}(t, \gamma^{\pm}(t))-\hat{\mu}(0, \gamma^{\pm}(0))
   & \le & -\frac{t}{2c_1 R_1}
   +d\int_{0}^t\Big(\frac{1}{2c_1 R_1}+\frac{M}{r_0^2}\Big)\,ds
   \\ & = & -\bigg(\frac{1-d}{2c_1 R_1}-d\,\frac{M}{r_0^2}\bigg)t
   \\ & \le & -\bigg(\frac{1}{4c_1 R_1}-d\,\frac{M}{r_0^2}\bigg)t
   \\ & \le & -\frac{t}{8c_1 R_1},\quad t\in [0, T[.
\end{eqnarray*}
Hence Lemma \ref{hatmu-lem}(a) leads to the estimate
\begin{eqnarray}\label{2step}
   |\gamma^{\pm}(t)-\gamma^{\pm}(0)|
   & = & \bigg|\int_{0}^t\,e^{(\mu-\lambda)(s,\,\gamma^{\pm}(s))}\,ds\bigg|
   \le\int_{0}^t\,e^{\hat{\mu}(s,\,\gamma^{\pm}(s))}\,ds
   \nonumber\\ & \le & e^{\hat{\mu}(0,\,\gamma^{\pm}(0))}
   \int_{0}^t\,e^{-\frac{s}{8c_1 R_1}}\,ds\le 8c_1 R_1 \sqrt{\frac{R_1-r_1}{R_1}},
\nonumber
\end{eqnarray}
where we used (\ref{emuhat}) in the last inequality.
By the third condition in (\ref{smallstrip}),
\[
\sqrt{\frac{R_1-r_1}{R_1}}<\frac{r_1-r_0}{16 c_1R_1},
\]
so that
\[
|\gamma^{\pm}(t)-\gamma^{\pm}(0)|<\frac{r_1-r_0}{2},\quad t\in [0, T[.
\]
Since $\gamma^-(0)-\gamma^+(0)=r_1-r_0$, this implies that
$\gamma^-(T)-\gamma^+(T) > 0$. Hence, if we choose $T$ in (\ref{fg-esti})
to be maximal, then $T=\infty$, i.e., $\gamma^+$ and $\gamma^-$ do never intersect.
This completes the proof of Theorem \ref{genmat}.
\prfe
\smallskip

\noindent
{\bf Remark.} 
In the above proof we have obtained the
more explicit information that
\begin{equation} \label{precise}
\lim_{s\to \infty}\gamma^+(s) < \frac{r_0+r_1}{2},
\quad m(s,\gamma^+(s))= M-M_\mathrm{out},\; s\geq 0,
\end{equation}
the latter since all the matter originally to the right of
$\gamma^- (s) > \gamma^+(s)$ necessarily stays there.

\section{Proof of Theorem~\ref{vlasov2}} \label{secvlasov2}
\setcounter{equation}{0}
%
%
We first check that the (DEC),
(NNP), and (GLO) conditions hold for Vlasov matter.
Then we show that there exists a class of initial data such that
the corresponding solutions satisfy the (GCC) condition with $c_1=3$.
Hence Theorem \ref{vlasov2} will follow from Theorem \ref{genmat}.

The characteristic system associated to the Vlasov equation (\ref{vlasov}) is
\begin{eqnarray}
   \frac{dR}{ds} & = & e^{(\mu-\lambda)(s,\,R)}\,\frac{W}{E}, \label{char1}
   \\[1ex] \frac{dW}{ds} & = & -\lambda_t(s, R)W-e^{(\mu-\lambda)(s,\,R)}\mu_r(s, R)E
   +e^{(\mu-\lambda)(s,\,R)}\frac{L}{R^3 E}, \label{char2}
   \\[1ex] \frac{dL}{ds} & = & 0. \label{char3}
\end{eqnarray}
If $s\mapsto (R, W, L)(s)$ is a solution with data $(R, W, L)(0)=(r, w, L)$, then
\[
f(s, R(s), W(s), L)=\open{f}(r,w, L)
\]
is constant in $s$. Hence $(R(s), W(s), L)\in \supp f(s)$ iff
$(r,w,L) \in \supp \open{f}$.
Such characteristics will be addressed as characteristics in
$\supp f$.

Direct inspection of the definition in (\ref{p})
shows that (NNP) holds for Vlasov matter.
It is moreover well-known that the (DEC) condition is satisfied for Vlasov matter;
see \cite[Sec.~1.4]{And05}. Alternatively, we can check
(\ref{nnrho}) and (\ref{mattergeodesic}) directly. The latter follows from
(\ref{char1}) above, whereas the former is a consequence of the expressions
for the matter terms given in (\ref{rho}), (\ref{p}), and (\ref{j}).

To see that the
(GLO) condition holds for any regular initial data set
we argue as follows. First of all, a regular initial data launches
a local-in-time solution on some time interval $[0,T[$, and the
corresponding theorems in \cite{RR1} or \cite{Rein95} also give
a condition under which this local solution can be extended to
a global one. In order to see that the local solution can always
be extended to the whole outer domain $D$ we first observe that
the spherically symmetric Einstein-Vlasov system on $D$,
with (\ref{e2lamb}) and (\ref{m-def}) replacing the usual boundary condition
of a regular centre and with (\ref{gamma+}) included,
has again a well-posed initial value problem for regular
data supported in $]r_0,\infty[$. This can be shown in the same
way as for the system on the whole space, the essential point being that
no characteristic of the Vlasov equation can enter region $D$
at the boundary $r=\gamma^+(t)$. To the local solution on $D$ we can now apply
the arguments from \cite{RRS} and conclude that the solution
exists on all of $D$. This is possible due to the fact that
the estimates in \cite{RRS} address a situation where matter is bounded away
from the centre or is controlled in a neighborhood of the centre
so that these estimates can be applied on $D$.
We emphasize that for our present analysis only the behaviour
of the solution on $D$ plays a role. We have chosen to present
our results in the form that we have Vlasov matter also inside
$r<\gamma^+(t)$, and this part of the solution may or may not break down,
but this is irrelevant for our arguments.

Hence it remains to show that the (GCC) condition holds.
To this end we let $0<r_0<r_1<R_1$, $R_0=(r_1+R_1)/2$, and $M=r_1/2$.
For a parameter $W_-<0$ to be specified
below and regular data $\fn$ with ADM mass $M$ we formulate the following

\smallskip

\noindent
{\bf General support condition:} For all $(r,w,L) \in \supp \fn\,$ the following holds:
\[
r \in ]0,r_0] \cup [R_0,R_1],
\]
and if $r\in [R_0,R_1]$ then
\[
w \leq W_-
\]
and also
\begin{equation}\label{hypoL}
   0< L <\frac{3L}{\eta}\,\open{m}(\eta) +\eta\,\open{m}(\eta),\ \eta\in [r_0,R_1].
\end{equation}
We use the notation $\open{m}$ when $\rho=\open{\rho}\,$ in (\ref{m-def}).
Furthermore, we abbreviate
\begin{equation} \label{Gammadef}
\Gamma = \Gamma(r_1,R_1) := \sqrt{\frac{R_1-r_1}{R_1+r_1}}.
\end{equation}
The following lemma shows that if the support condition holds,
then the particles in the outer domain $D$ keep moving inward
in a controlled way.
\begin{lemma}\label{ingoinglemma}
Let $\fn$ be regular and satisfy the general support condition
for some  $W_-<0$.
Then for all $(r,w,L) \in \supp f(t)$ such that $(t, r)\in D$,
\[
w \leq  \Gamma(r_1,R_1) W_-.
\]
In particular, $j\leq 0$ on $D$.
\end{lemma}
\textbf{Proof\,:} Let $[0, T[$ denote the maximal time
interval such that for $t < T$
\begin{equation}\label{bootcd}
w < 0 \ \mbox{for}\ (r,w,L)\in\supp f(t)\ \mbox{with}\
(t,r) \in D.
\end{equation}
Since $W_-<0$, $T>0$ by continuity.
By the definition of $j$,
\begin{equation}\label{jle0}
j(t, r)\le 0 \ \mbox{for}\ (t,r)\in D_T:=D \cap ([0,T[\times [0,\infty[).
\end{equation}
Let $(R, W, L)(s)$ be a characteristic in $\supp f$.
Then
\begin{eqnarray*}
   \frac{d}{ds}(e^{-\lambda}W) & = &
   -\,e^{-\lambda}\Big(W\lambda_t+W\lambda_r\frac{dR}{ds}-\frac{dW}{ds}\Big)
   \\ & = & \frac{4\pi R}{E}\,e^{\mu}(2WEj-W^2\rho-E^2 p)
   +e^{\mu}\Big(1-\frac{2m}{R}\Big)\frac{L}{R^3 E}
   \\ & & +\,e^{\mu}\frac{m}{R^2}\,\Big(\frac{w^2}{E}-E\Big)
   \\ & = & -\,\frac{4\pi^2}{R}\,e^{\mu}\,\int_{-\infty}^\infty\int_0^\infty
   \bigg[\sqrt{\frac{\tilde{E}}{E}}\,w
   -\sqrt{\frac{E}{\tilde{E}}}\,\tilde{w}\bigg]^2\,f\,d\tilde{L}\,d\tilde{w}
   \\ & & -\,e^{\mu}\frac{m}{R^2}\bigg(\frac{1+L/R^2}{E}+\frac{2L}{R^2 E}\bigg)
   +e^{\mu}\frac{L}{R^3 E}\,,
\end{eqnarray*}
where $E=E(R, W, L)$ and $\tilde{E}=\tilde{E}(R, \tilde{w},
\tilde{L})$. Therefore
\[
\frac{d}{ds}(e^{-\lambda}W)\le -e^{\mu}\frac{m}{R^2}\bigg(\frac{1+L/R^2}{E}
   +\frac{2L}{R^2 E}\bigg)+e^{\mu}\frac{L}{R^3 E}.
\]
Differentiating (\ref{e2lamb}) w.r.t.\ $t$ and using (\ref{ein3})
leads to $m_t=-4\pi r^2 e^{\mu-\lambda} j$,
which by (\ref{jle0}) is non-negative on $D_T$.
It follows that $m(s, r)\ge m(0, r)=\open{m}\,(r)$.
Thus as long as the characteristic remains in $D_T$,
\begin{eqnarray*}
   \frac{d}{ds}(e^{-\lambda}W)
   & \le & -e^{\mu}\frac{\open{m}\,(R)}{R^2}\bigg(\frac{1+L/R^2}{E}
   +\frac{2L}{R^2 E}\bigg)+e^{\mu}\frac{L}{R^3 E}
   \\ & = & e^{\mu}\,\frac{1}{R^3 E}\bigg(L-\frac{3L}{R}\,\open{m}\,(R)
   -R\,\open{m}\,(R)\bigg).
\end{eqnarray*}
Now $R(0)\in [R_0, R_1]$ and $\dot{R}(s)\le 0$ by (\ref{char1})
and (\ref{bootcd}) yields $R_1\ge R(0)\ge R(s)\geq \gamma^+(s) \geq r_0$.
Hence condition (\ref{hypoL}) implies that,
as long as the characteristic remains in $D_T$,
$\frac{d}{ds}(e^{-\lambda}W)<0$, so that
\[
W(s)\le e^{\lambda(s,\,R(s))-\lambda(0,\,R(0))}\,W_- .
\]
But $\lambda\ge 0$, so $W_-<0$ leads to
\[
W(s)\le\Big(\min_{r\in [R_0, R_1]} e^{-\lambda(0,\,r)}\Big)\,W_-.
\]
In view of (\ref{e2lamb}),
\[
e^{-\lambda(0,\,r)}\geq \sqrt{1-\frac{2M}{R_0}}=\sqrt{\frac{R_1-r_1}{R_1+r_1}},
\quad r\in [R_0,R_1],
\]
and recalling (\ref{Gammadef}) it follows that
\[
W(s)\leq \Gamma(r_1,R_1) W_-<0
\]
as long as the characteristic remains in $D_T$.
By the maximality of $T$ in (\ref{bootcd}), $T=\infty$,
and the proof is complete.
\prfe

In order to specify the initial data set ${\cal{I}}_2$, let
$r_0,\, r_1,\, M$, and $M_\mathrm{out}$ be given as in Section~\ref{secvlasres}
and let $R_1$ be such that (\ref{smallstrip}) holds for $\kappa=6$.
We require that $W_-<0$ satisfies the estimate
\begin{equation}\label{condw}
   \Gamma(r_1,R_1)\, |W_-|\geq 1.
\end{equation}
Then
\begin{eqnarray}\label{I2def}
   {\cal{I}}_2 := \Bigl\{ \fn
   &\mid&
   \fn \ \mbox{is regular, satisfies (\ref{checkM}), (\ref{M-checkM}),
   the general support condition,}\nonumber \\
   &&
   \mbox{and for}\ (r,w,L)\in \supp \fn\ \mbox{with}\ r\in [R_0,R_1],
   \sqrt{L}/r_0 \leq  \Gamma\, |W_-|\Bigr\}.\nonumber \\
   &&
\ \label{condL1}
\end{eqnarray}
Consider now a solution $f$ launched by initial data
from this set. Condition (\ref{condw}) and Lemma \ref{ingoinglemma}
imply that
\begin{equation}\label{condw2}
   |w|\geq\Gamma(r_1,R_1)\,|W_-|\geq 1\quad\mbox{on}\quad\supp f \cap D,
\end{equation}
and since $L$ is conserved along characteristics, (\ref{condL1})
leads to $\sqrt{L}/r \leq \sqrt{L}/r_0 \leq |w|$ for all
particles in $\supp f \cap D$. Hence the definition (\ref{rho}) of $\rho$
implies that on $D$,
\begin{eqnarray}
\rho(t,r)
&\leq&
\frac{\pi}{r^2}\int_{-\infty}^{\infty}\int_0^{\infty} f\,dL\,dw
+ \frac{\pi}{r^2}\int_{-\infty}^{\infty}\int_0^{\infty} |w|f\,dL\,dw\nonumber\\
&&
{}+
\frac{\pi}{r^2}\int_{-\infty}^{\infty}\int_0^{\infty} \sqrt{L}/r f\,dL\,dw\nonumber\\
&\leq&
3\,\frac{\pi}{r^2}\int_{-\infty}^{\infty}\int_0^{\infty} |w|f\,dL\,dw =3\,|j(t,r)|.
\end{eqnarray}
Accordingly, ${\cal{I}}_2$ satisfies the (GCC) condition with $c_1=3$,
and Theorem~\ref{vlasov2} follows from Theorem~\ref{genmat}. \prfe

We briefly show that the set ${\cal{I}}_2$ is far from empty.
Therefore fix $0<r_0<r_1<R_0<R_1$, $M=r_1/2$, and $0<M_\mathrm{out}<M$
such that $R_0=(r_1+R_1)/2$, (\ref{icnts}), and (\ref{smallstrip})
are satisfied. Let $0\leq f_1\in C^1$ have $r$-support in $[r_0-\delta,r_0]$
for some $0<\delta<r_0/9$, and let $0\leq f_2\in C^1$ have $r$-support in $[R_0, R_1]$.
Fix the compact $w$-support of $f_2$ in $]-\infty,W_{-}]$ with
$W_-<0$ such that (\ref{condw}) holds, and fix its $L$-support in $[0, L_2]$ so that
\[
\frac{\sqrt{L_2}}{r_0} \leq \Gamma(r_1,R_1)\,|W_-|
\]
and
\[
L <(M-M_\mathrm{out})\Big(\frac{3L}{\eta}+\eta\Big),\quad L\in[0, L_2],
\quad\eta\in [r_0,R_1].
\]
Now take $\open{f}=A f_1+B f_2,$ where $A>0$ and $B>0$ are chosen
such that (\ref{checkM}) and (\ref{M-checkM}) are satisfied.
Note that $\open{m}(\eta)\ge M-M_\mathrm{out}$ for $\eta\in [r_0, R_1]$,
whence (\ref{hypoL}) holds as well; thus the general support condition
if verified. It remains to check (\ref{notsinit2}). If $r\in ]0, r_0-\delta]$,
then $\open{m}(r)=0$. If $r\in [r_0-\delta, R_0]$, then $\open{m}(r)\le M-M_\mathrm{out}$
yields in view of (\ref{icnts}),
\[
\frac{2\open{m}}{r}\leq \frac{2(M-M_\mathrm{out})}{r_0-\delta}<1.
\]
If $r\in [R_0, \infty[$, then
\[
\frac{2\open{m}}{r}\leq \frac{2M}{R_0}<1,
\]
since $2M=r_1<R_0.$ Hence $\open{f}$ is regular
and has all the properties that are required
in the definition of ${\cal{I}}_2$.
\smallskip

\noindent
{\bf Remark.} The set ${\cal{I}}_2$ has ``non-empty interior'',
in the sense that sufficiently small perturbations
of initial data in the ``interior'' of this set belong to ${\cal I}_2$ as well,
provided that the support is changed very little and $M$ is left invariant.
This is due to the fact that the various parameters entering into the definition
of ${\cal{I}}_2$ are defined in terms of inequalities and hence can be varied.

%
%
\section{Proof of Theorem~\ref{vlasov1}} \label{secvlasov1}
\setcounter{equation}{0}
%
%
The set up is closely related to the set up in the proof
of Theorem~\ref{vlasov2}.
As we saw above, the (DEC), (NNP), and (GLO) conditions are
satisfied for Vlasov matter, and we will again construct an initial data
set such that the (GCC) condition holds with $c_1=3$. However, since this result
relies on condition (\ref{mediumstrip}) instead of (\ref{smallstrip}),
we cannot simply invoke Theorem \ref{genmat} after the (GCC) condition has been verified;
instead an additional step needs to be added to the proof. For this new argument
a slightly stronger condition on the momentum variable $w$ needs to be imposed
on $\supp \fn$. We now require that $W_-<0$ satisfies
\begin{equation}\label{condwt1}
   \Gamma(r_1,R_1)^2 |W_-|^2\geq \frac{10}{d},
\end{equation}
where
\[
d:=\min\left\{\frac12,\frac{r_0}{12 R_1},\frac{r_1-r_0}{300 R_1}\right\}.
\]
Then
\begin{eqnarray}\label{I1def}
{\cal{I}}_1 := \Bigl\{ \fn
&\mid&
\fn \ \mbox{is regular, satisfies (\ref{checkM}), (\ref{M-checkM}),
the general support condition,}\nonumber \\
&&
\mbox{and for}\ (r,w,L)\in \supp \fn\ \mbox{with}\ r\in [R_0,R_1],
\sqrt{L}/r_0 \leq 1.
\Bigr\}
\ \label{condL1t1}
\end{eqnarray}
The same construction as at the end of the previous section
shows that this set is not empty, and the same remark as at the end of the
previous section applies.

Let $f$ be a solution launched by initial data from ${\cal{I}}_1$.
It is clear from these conditions that Lemma $\ref{ingoinglemma}$ applies,
and since $10/d\geq 1$, it follows that (\ref{condw2}) holds as well.
Thus the argument leading to $\rho\leq 3|j|$ on $D$
in the proof of Theorem~\ref{vlasov2} applies again.
Hence, the (GCC) condition is satisfied with $c_1=3$.

Consider the expression
\[
\rho(s, r)-p(s, r)=
\frac{\pi}{r^2}\int_{-\infty}^\infty \int_0^\infty
\Big(E-\frac{w^2}{E}\Big) \,f(s, r, w,L)\,dL\,dw.
\]
Since $E^2\ge w^2\ge\Gamma^2(r_1,R_1)\,W_-^2$ by Lemma \ref{ingoinglemma},
we get for $r\in [\gamma^+(s), R_1]$ from $\sqrt{L}/r_0\le 1$,
\begin{equation}\label{c0-def}
   E-\frac{w^2}{E}=\frac{1}{E}\,(E^2-w^2)=\frac{1}{E}\,\Big(1+\frac{L}{r^2}\Big)
   \le\frac{2}{E}\le\frac{2}{\Gamma^2\,W_-^2}\,E=:c_0 E,
\end{equation}
so that
\begin{equation}\label{c0}
   \rho(s, r)-p(s, r)\le c_0\rho(s, r).
\end{equation}
After this preparation, we again show that the out- and ingoing null geodesics
$\gamma^+$ and $\gamma^-$ do not intersect. We choose
$T\in ]0,\infty[$ such that (\ref{fg-esti}) holds.
In this case we cannot rely on the smallness of $e^{\hat{\mu}}$ as in the proof
of Theorem \ref{genmat}, so we need to control the evolution also when
$e^{\hat{\mu}}$ is not small.
For this part the estimate (\ref{c0}) is essential.

We fix $t_\ast^{\pm}\in [0, T[$ by requiring that
\[
e^{(\hat{\mu}+\lambda)(s,\,\gamma^{\pm}(s))} > d\,\,\mbox{for}\,\,
   s\in [0, t_\ast^{\pm}[,\quad
   e^{(\hat{\mu}+\lambda)(s,\,\gamma^{\pm}(s))}\le d\,\,\mbox{for}\,\,
   s\in [t_\ast^{\pm}, T[.
\]
First we note that $t_\ast^{\pm}$ is well-defined, since by Lemma \ref{gammapm}(c),
\begin{equation}\label{monotone}
   \frac{d}{ds}\,e^{(\hat{\mu}+\lambda)(s,\,\gamma^{\pm}(s))}\le 0.
\end{equation}
{\em Step 1:}
Consider $s\in [0, t_\ast^{\pm}]$; if $t_\ast^{\pm}=0$,
then this step is omitted. For $\eta\ge\gamma^{\pm}(s)$,
\[
d\le e^{(\hat{\mu}+\lambda)(s,\,\gamma^{\pm}(s))}
   \le e^{(\hat{\mu}+\lambda)(s,\,\eta)},
\]
since $(\hat{\mu}+\lambda)_r=4\pi r\rho\,e^{2\lambda}\ge 0$ by (\ref{gsto}).
Hence Lemma \ref{hatmu-lem}(c) and (\ref{c0}) yield
\begin{eqnarray*}
   (\mu-\lambda)(s, \gamma^{\pm}(s)) & = & 2\hat{\mu}(s, \gamma^{\pm}(s))
   +\int_{\gamma^{\pm}(s)}^\infty 4\pi\eta\,(\rho-p)(s, \eta)\,e^{2\lambda(s,\,\eta)}\,d\eta
   \\ & \le & 2\hat{\mu}(s, \gamma^{\pm}(s))
   +\frac{c_0}{d}\int_{\gamma^{\pm}(s)}^\infty 4\pi\eta\,\rho(s, \eta)
   \,e^{(\hat{\mu}+\lambda)(s,\,\eta)}e^{2\lambda(s,\,\eta)}\,d\eta
   \\ & \le & 2\hat{\mu}(s, \gamma^{\pm}(s))+\frac{c_0}{d},
\end{eqnarray*}
where for the last estimate Lemma \ref{int0infty} has been used.

Now we make the following observation: There is at least one characteristic
$(\bar{R},\bar{W},\bar{L})(s)$ with $\bar{R}(0)\in [R_0,R_1],$
which does not leave the strip $[r_1,R_1]$ during the finite time interval $[0, T]$.
In fact, if at time $t=T$ all characteristics had left the strip $[r_1, R_1]$
(and thus had entered the region $r<r_1$), then $m(T, r_1)=M$.
From (\ref{e2lamb}) and $2M=r_1$ it would follow that $\lambda(T,r_1)=\infty$.
However, this contradicts the (GLO) condition which holds for Vlasov matter.

Since $\gamma^{\pm}(s)\le r_1\le \bar{R}(s)$, and since $\hat{\mu}_r\geq 0$,
we thus obtain in view of Lemma \ref{hatmu-lem}(a) that
\begin{eqnarray*}
   (\mu-\lambda)(s, \gamma^{\pm}(s))
   & \le & 2\hat{\mu}(s,\gamma^{\pm}(s))
   +\frac{c_0}{d}\le 2\hat{\mu}(s, \bar{R}(s))+\frac{c_0}{d}
   \\ & \le & (\mu-\lambda)(s, \bar{R}(s))+\frac{c_0}{d},
   \quad s\in [0, t_\ast^{\pm}].
\end{eqnarray*}
Next note that $|W|\ge 1$ by (\ref{condw2}),
and hence due to (\ref{char1}) and observing
$\bar{R}^2\ge r_0^2\ge L$,
\[ |\dot{\bar R}|=\frac{|W|}{E}\,e^{\mu-\lambda}
   \geq\frac{|W|}{\sqrt{2+W^2}}\,e^{\mu -\lambda}
   \geq\frac{1}{2}\,e^{\mu-\lambda}. \]
Therefore for all $t\in [0, t_\ast^{\pm}]$ the estimate
\begin{eqnarray}\label{1step}
|\gamma^{\pm}(t)-\gamma^{\pm}(0)|
& = &
\bigg|\int_0^t\pm\,e^{(\mu-\lambda)(s,\,\gamma^{\pm}(s))}\,ds\bigg|
\le e^{\frac{c_0}{d}}\int_0^t e^{(\mu-\lambda)(s,\,\bar{R}(s))}\,ds\nonumber\\
& \le &
-2e^{\frac{c_0}{d}}\int_0^t\dot{\bar{R}}(s)\,ds
=2e^{\frac{c_0}{d}}(\bar{R}(0)-\bar{R}(t))\nonumber\\
& \le & 2e^{\frac{c_0}{d}}(R_1-r_1)
\end{eqnarray}
is obtained. \\
{\em Step 2:}  Let $t\in [t_\ast^{\pm}, T[$;
if $t_\ast^{\pm}=T$, then this step is omitted.
The arguments here are basically the ones
presented in Section~\ref{secgenmatproof}.
The computation leading to (\ref{est}) is almost identical, and
\begin{eqnarray}\label{est2}
&&
\hat{\mu}(t, \gamma^{\pm}(t))-\hat{\mu}(t_{\ast}^{\pm},
   \gamma^{\pm}(t_{\ast}^{\pm}))
\nonumber\\
&&
\qquad \le  -\frac{t-t_{\ast}^{\pm}}{2c_1 R_1}
   +\int_{t_{\ast}^{\pm}}^t\Big(\frac{1}{2c_1 R_1}
   +\frac{m(s, \gamma^{\pm}(s))}{\gamma^{\pm}(s)^2}\Big)\,
   e^{(\mu+\lambda)(s,\,\gamma^{\pm}(s))}\,ds
\end{eqnarray}
for $c_1=3$. By Lemma \ref{hatmu-lem}(b), $e^{(\mu+\lambda)(s,\,\gamma^{\pm}(s))}
\le e^{(\hat{\mu}+\lambda)(s,\,\gamma^{\pm}(s))}\le d$.
Next we use the facts that $m/r<1/2,\; \gamma^{\pm}(s)\ge r_0$,
and the definition of $d$ to obtain the estimate
\begin{eqnarray*}
   \hat{\mu}(t, \gamma^{\pm}(t))-\hat{\mu}(t_\ast^{\pm}, \gamma^{\pm}(t_\ast^{\pm}))
   & \le & -\frac{1}{2c_1 R_1}(t-t_\ast^{\pm})
   +d\int_{t_\ast^{\pm}}^t\Big(\frac{1}{2c_1 R_1}+\frac{1}{2r_0}\Big)\,ds
   \\ & = & -\bigg(\frac{1-d}{2c_1 R_1}-d\,\frac{1}{2r_0}\bigg)(t-t_\ast^{\pm})
   \\ & \le & -\bigg(\frac{1}{4c_1 R_1}-d\,\frac{1}{2r_0}\bigg)(t-t_\ast^{\pm})
   \\ & \le & -\frac{1}{8c_1 R_1}(t-t_\ast^{\pm}),\quad t\in [t_\ast^{\pm}, T[.
\end{eqnarray*}
Hence by Lemma \ref{hatmu-lem}(a),
\begin{eqnarray}\label{22step}
   |\gamma^{\pm}(t)-\gamma^{\pm}(t_\ast^{\pm})|
   & = & \bigg|\int_{t_\ast^{\pm}}^t\,e^{(\mu-\lambda)(s,\,\gamma^{\pm}(s))}\,ds\bigg|
   \le\int_{t_\ast^{\pm}}^t\,e^{\hat{\mu}(s,\,\gamma^{\pm}(s))}\,ds
   \nonumber\\ & \le & e^{\hat{\mu}(t_\ast^{\pm},\,\gamma^{\pm}(t_\ast^{\pm}))}
   \int_{t_\ast^{\pm}}^t\,e^{-\frac{(s-t_\ast^{\pm})}{8c_1 R_1}}\,ds
   \nonumber\\ & \le & e^{(\hat{\mu}+\lambda)(t_\ast^{\pm},\,\gamma^{\pm}(t_\ast^{\pm}))}
   \int_{t_\ast^{\pm}}^\infty\,e^{-\frac{(s-t_\ast^{\pm})}{8c_1 R_1}}\,ds\le 8c_1 R_1 d.
\end{eqnarray}
Adding the contributions (\ref{1step}) from Step~1 and (\ref{22step}) from Step~2,
the final estimate
\[
|\gamma^{\pm}(t)-\gamma^{\pm}(0)|\le 2e^{c_0/d}(R_1-r_1)
+8c_1 R_1 d
\]
is obtained for all $t\in [0, T[$. From (\ref{c0-def}) and (\ref{condwt1})
we have $c_0/d\leq 1/5$. The third condition on $d$
together with (\ref{mediumstrip}) thus imply that
\[
|\gamma^{\pm}(t)-\gamma^{\pm}(0)|<\frac{r_1-r_0}{2}.
\]
As in the proof of Theorem \ref{genmat} we
conclude that $\gamma^+$ and $\gamma^-$ do not intersect,
completing the proof of Theorem \ref{vlasov1}.
\prfe

\smallskip

\noindent
{\bf Remarks.}
(a)
The sharper estimates stated in (\ref{precise})
clearly hold also in this case.\\
(b)
The solution must necessarily enter the regime
of Step~2, more precisely,
\[
\lim_{s\to \infty} e^{(\hat{\mu}+\lambda)(s,\,\gamma^{\pm}(s))} = 0
\]
for both null geodesics. Otherwise, the monotonicity implied
by Eqn.~(\ref{monotone}) yields a positive constant $c>0$ such
that $e^{(\hat{\mu}+\lambda)(s,\,\gamma^{\pm}(s))}>c$ for all time,
and hence,
\[
|\dot \gamma^\pm| = e^{\mu - \lambda} = 
e^{\hat{\mu} + \lambda}e^{\check{\mu} -  2\lambda}
> c e^{\check{\mu} -2\lambda}.
\]
Since no matter can cross the two null geodesics,
\begin{eqnarray*}
(\check{\mu} -2\lambda)(s,r)
&=&
\int_r^\infty 4 \pi \eta(2 \rho - p)e^{2\lambda} d\eta + 2\hat{\mu}(s,r)\\
&\geq&
2\hat{\mu}(s,r) = -2 \int_r^\infty \frac{\open{m}(r_0)}{\eta^2}
\frac{1}{1-2\open{m}(r_0)/\eta}\,d\eta\\
&=&
\ln \frac{r-2\open{m}(r_0)}{r}
\end{eqnarray*}
for $r=\gamma^\pm(s)$. If we insert this into the estimate for 
$\dot \gamma^\pm$ it follows that this quantity is bounded from below
by a positive constant which contradicts the finite limits of
$\gamma^\pm(s)$ as $s\to \infty$.

It remains to prove Cor.~\ref{ssinthemiddle}.

\noindent
{\bf Proof of Corollary~\ref{ssinthemiddle}\,:} 
Let $f_s$ be a static solution. By \cite{And2},
$2m_s(r)/r < 8/9$ for $r>0$ where 
$m_s$ is the local ADM mass 
induced by $f_s$. In particular, $M_s < r_s/2 < r_1/2 = M$, and (\ref{icnts})
holds. As described above we can now specify the matter distribution 
for $r\geq r_0$, and we obtain
initial data $\open{f}$ in ${\cal{I}}_1$ or
in ${\cal{I}}_2$ which coincide with the
given static solution for $0\leq r\leq r_0$. 

Since no matter travels from the outer domain $D$ to the inner
one where $r\leq \gamma^+(t)$, the only way the matter in the outer
domain can affect the static solution is through the metric.
Consider the time-independent
version of the Vlasov equation (\ref{vlasov}). Dropping all the
time derivatives we see that in the remaining equation the
factor $e^{\lambda - \mu}$ can be canceled. Therefore, the static
Einstein-Vlasov system is formulated in terms the quantities $f, \lambda$,
and $\mu_r$,
but not $\mu$ itself. By (\ref{e2lamb}) and (\ref{mur}),
$\lambda$ and $\mu_r$ are on $r\leq \gamma^+(t)$
not affected by the matter in the outer domain $D$,
and therefore $f=f_s, \lambda, \mu_r$ remain time-independent 
for $r\leq \gamma^+(t)$.
\prfe 

Notice that the metric coefficient $\mu$ of course does change on the interior
region, cf.\ Thm~\ref{bh} (b).

%
%
\section{Proof of Theorem~\ref{bh}}\label{bhproof}
\setcounter{equation}{0}
%
%
As a first step we estimate $\mu -\lambda$ from below for $r > 2 M$, using
Lemma~\ref{hatmu-lem}~(a):
\beas
(\mu-\lambda)(t,r)
&\geq&
2\hat \mu (t,r) = -2 \int_r^\infty \frac{m(t,\eta)}{\eta^2}
e^{2\lambda(t,\eta)}d\eta\\
&=&
-2 \int_r^\infty \frac{m(t,\eta)}{\eta\,(\eta - 2 m(t,\eta))}d\eta
\geq
-2 \int_r^\infty \frac{M}{\eta\,(\eta - 2 M)}d\eta \\
&=&
\ln\frac{r-2M}{r},\ r> 2 M.
\eeas
Now consider any characteristic in the matter support, and let $R(t)$ denote
its radial coordinate. Then by Lemma~\ref{ingoinglemma} and as long as $R(t) > 2 M$,
\[
\frac{dR}{ds} = e^{(\mu-\lambda)(s,R)}\frac{W}{E} \leq - C e^{(\mu-\lambda)(s,R)}
\leq - C \frac{R-2M}{R};
\]
for initial data from the set ${\cal I}_1$ respectively ${\cal I}_2$
one can take $C:=\Gamma |W_-|/\sqrt{2+\Gamma^2 W_-^2}$ respectively $C:=1/\sqrt{3}$.
Integrating this differential inequality we find that as long as  $R(t) > 2 M$
the estimate
\beas
-C t 
&\geq&
 \int_{R(0)}^{R(t)} \frac{r}{r-2M} dr = R(t) - R(0) + 
2 M \ln \frac{R(t)-2M}{R(0) - 2M}\\
&\geq&
2 M - R_1 + 2 M \ln \frac{R(t)-2M}{R(0) - 2M}
\eeas
holds, and hence
\[
R(t) \leq 2 M  + (R_1 - 2 M)e^{\frac{1}{2 M}(R_1 - 2 M - C t)},
\]
which proves the support estimate in part (a).
Since all the matter, which has ADM mass $M$, is contained
in the region where
$r \leq 2 M + \alpha e^{-\beta t}=: \sigma(t)$, the assertion on the metric follows.
Moreover, for any $r\leq \sigma(t)$ the monotonicity of $\mu$ with respect to $r$
implies that
\[
\mu(t,r) \leq \ \mu(t,\sigma(t)) = \hat\mu(t,\sigma(t)) =
\ln \left(\frac{\sigma(t) -2 M}{\sigma(t)}\right)^{1/2},
\]
which is the first assertion of part (b). The second follows immediately since 
the integral $\int_0^\infty e^{\mu(t,r)}dt$ is the proper length
of a coordinate line of constant $r,\theta$, and $\varphi$ in the outer region
$D$. This completes the proof of part (b). 

As to (c) we first observe that any radially outgoing null geodesic which
enters the region $r > 2 M$ escapes to $r=\infty$ and is future complete,
since by part (a) the metric on $r>2 M+ \epsilon$ where $\epsilon >0$ is arbitrary
eventually equals the Schwarzschild one for which the asserted
properties of the geodesics hold. Now consider the extremal geodesic
$\gamma^\ast$. If there existed some time $t>0$ such that $\gamma^\ast(t) > 2 M$,
then by continuous dependence on the initial data the same would be true
for all radially outgoing null geodesics with $\gamma(0)$ 
sufficiently close  to but less than $r^*$.
Hence such geodesics would escape to $r=\infty$ in contradiction to the
definition of $r^\ast$. This shows that the extremal,
radially outgoing null geodesic $\gamma^\ast$ has the property that 
$\lim_{t\to\infty} \gamma^\ast(t) \leq 2 M$. 

It remains to show that the limit above cannot be strictly less than
$2 M$. To this end we consider a radially outgoing null geodesic as long
as $\gamma(t) < \sigma(t) = 2 M + \alpha e^{-\beta t}$. Then
\[
\frac{d\gamma}{ds} =
e^{(\mu-\lambda)(s,\gamma(s))} \leq e^{\mu(s,\sigma(s))} 
=
\left(\frac{\sigma(s) -2 M}{\sigma(s)}\right)^{1/2}
\leq
C e^{-\beta s/2},
\]
and hence for any $0\leq t_0 \leq t$ and as long as
$\gamma(t) < \sigma(t)$, 
\[
\gamma(t) \leq \gamma(t_0) + C e^{-\beta t_0/2},
\]
where the constant $C>0$ again depends only on the initial data set.
Now assume that $R^\ast := \lim_{t\to \infty}\gamma^\ast(t) < 2 M$, 
choose $t_0>0$ such that 
$R^\ast + C e^{-\beta  t_0/2} < 2 M$, and consider the radially outgoing 
null geodesic $\gamma^{\ast\ast}$ with $\gamma^{\ast\ast}(t_0) = R^\ast$.
Then by construction,  $\gamma^{\ast\ast}(t) < 2 M < \sigma(t)$ for all $t\geq t_0$,
and since $\gamma^{\ast\ast}(t_0) = R^\ast > \gamma^{\ast}(t_0)$ it follows
that $\gamma^{\ast\ast}(0) > \gamma^{\ast}(0)=r^\ast$. Hence
$\gamma^{\ast\ast}$ is a radially outgoing null geodesic which at time 
$t=0$ starts to the right of $r^\ast$ and does not escape to $r=\infty$.
This is in contradiction to the definition of $r^\ast$. \prfe 

We conclude this section by proving the remark after Theorem~\ref{genmat}.
Under our general matter conditions the matter
is ingoing in the region $D$, in particular, the matter is for all time
restricted to the region where $r\leq R_1$. Hence for $r\geq R_1$
the metric is again equal to the Schwarzschild one with mass $M$,
and if we replace $2 M$ by $R_1$ in the above argument for part (c)
we obtain the assertions on $\gamma^\ast$ in the general matter
context. As to the divergence of $\mu$ we observe that
\[
\frac{d}{ds} \hat \mu (s,\gamma^-(s)) = 
\hat \mu_t (s,\gamma^-(s)) + \hat \mu_r(s,\gamma^-(s))\frac{d\gamma^-}{ds} (s) \leq 0
\]
so that the limit
$
\hat \mu_\infty := \lim_{s\to \infty}\hat \mu (s,\gamma^-(s))
$
exists. The fact that $\gamma^-(s)>0$ is decreasing with
\[
\left| \frac{d\gamma^-}{ds} (s)\right| = e^{(\mu-\lambda)(s,\gamma^-(s))}
\geq e^{2 \hat\mu (s,\gamma^-(s))} \geq e^{2 \hat\mu_\infty},
\]
implies that $\hat\mu_\infty = -\infty$. Since $\mu \leq \hat\mu$ we conclude
that
\[
\lim_{s\to \infty}\mu (s,\gamma^-(s)) = -\infty,
\] 
from which the assertion follows by the monotonicity of $\mu$
with respect to $r$.

\noindent
{\bf Acknowledgement\,:} The authors are grateful for discussions with A.~Rendall.

\end{document}